\documentclass[]{aa}
\usepackage{graphics}
\usepackage{amsmath}
\usepackage{epsfig}
 \begin{document}

 \title{Line emission from circumstellar disks around A stars\thanks{Based
on observations with ISO, an ESA project with instruments funded by ESA
Member States (especially the PI countries: France, Germany, The Netherlands
and the United Kingdom) and with the participation of ISAS and NASA.}}
 \titlerunning{Line emission from disks around A stars}

 \author{Inga~Kamp \inst{1}, Gerd-Jan~van~Zadelhoff \inst{1}, Ewine~F.~van~Dishoeck \inst{1}, Ronald~Stark \inst{2}}
 \authorrunning{Kamp et al.}
 \offprints{Inga~Kamp}
 \mail{kamp@strw.leidenuniv.nl}
 \institute{Leiden Observatory, PO Box 9513, NL-2300 RA Leiden, 
            The Netherlands \and Max-Planck-Institut f\"{u}r Radioastronomie,   
            Auf dem H\"{u}gel 69, D-53121 Bonn, Germany }
 \date{Received / Accepted}

 \abstract{ The nature of the tenuous disks around A stars has raised
considerable controversy in the literature during the past decade.
The debate whether or not the disk around $\beta$~Pictoris contains
gaseous molecular hydrogen is only the most recent example.  Since CO
is in general a poor tracer for the gas content of these low mass
disks, we discuss here detailed emission line calculations for
alternative tracers like C and C$^+$, based on recent optically thin
disk models by Kamp \& van Zadelhoff~(\cite{Kamp2}). The [\ion{C}{ii}] 
157.7~$\mu$m line was searched toward Vega and $\beta$~Pictoris 
--- the most prominent A stars with disks --- using ISO LWS 
data, and a tentative detection is reported toward the latter object. From 
a comparison with emission line observations as well as absorption line 
studies of both stars, the gas-to-dust ratio is constrained to lie between 
0.5 and 9 for $\beta$~Pictoris. For Vega the [\ion{C}{ii}] observations indicate
an upper limit of 0.2~M$_\oplus$ for the disk gas mass. Predicted line 
intensities of C$^+$ and C are presented for a range of models and appear 
promising species to trace the gas content in the disks around A stars with 
future instrumental capabilities (SOFIA, Herschel, APEX and ALMA). Searches
for CO emission should focus on the $J$=3--2 line.   }

 \maketitle

\section{Introduction}

A large number (20 -- 40 \%) of nearby A stars is surrounded by dust
disks (Cheng et al. \cite{cheng}, Habing et al. \cite{habingb}), see
for example $\beta$~Pictoris (20~Myr; Barrado y Navascu\'{e}s et
al. \cite{Barrado}) or HR4796A (8 Myr; Stauffer et
al. \cite{Stauffer}).  However, little is known about the nature of
these disks: are they dusty debris disks or do they still have dust
{\em and} gas and are possibly in a protoplanetary phase?

The disks around A stars have been extensively searched for gas. CO
rotational line observations revealed too little gas compared to the
detected dust. Upper limits suggest that the gas mass in these disks
can be up to a factor of 1000 lower than deduced from the typical
interstellar gas-to-dust ratio of 100 (Yamashita et
al. \cite{Yamashita}; Savoldini \& Galletta \cite{Savoldini};
Zuckerman et al.~\cite{Zuckerman}; Dent et al.~\cite{Dent}; Liseau \&
Artymowicz \cite{Liseau1}; Liseau \cite{Liseau2}). This has been
interpreted that the gas dissipates very rapidly from such disks on
timescales of a few million years.

In a previous paper, Kamp \& Bertoldi (\cite{Kamp1}) showed that in
low mass optically thin circumstellar disks CO is likely to be
photo-dissociated either by the stellar or by the interstellar
ultraviolet (UV) radiation. If effective shielding prevents CO from
being photo-dissociated, it is mainly found in the form of CO ice
mantles on the dust grains in the cool outer disk regions. This indicates
that CO is actually a poor tracer of the gas and one should look for
alternative tracers. A similar conclusion has been reached for
younger, more massive disks by Thi et al.~(\cite{Thib}).

In order to calculate the line emission from different atoms and
molecules in these disks a realistic modeling of the gas
temperature is needed. With the low densities found in these disks,
gas and dust are not in collisional equilibrium; hence the gas and
dust temperatures divert. The computation of the gas temperature was
performed by Kamp \& van Zadelhoff (\cite{Kamp2}), including the
relevant heating and cooling processes for the gaseous species.

Recently, Thi et al.~(\cite{Thi}) found hints of molecular hydrogen in
the disk around $\beta$~Pictoris using the Short Wavelength
Spectrometer (SWS) on board of the Infrared Space Observatory
(ISO). They tentatively detected the S(0) and S(1) lines and deduced
from the line ratio an excitation temperature of $\sim$100~K assuming
LTE ($n_{\rm cr} > 10^3$~cm$^{-3}$). The resulting H$_2$ mass is
$\sim$54~M$_\oplus$ with an uncertainty up to a factor of 3. On the
other hand, Lecavelier des Etangs et al.~(\cite{Lecavelier}) failed to
detect H$_2$ line absorption superposed on the broad \ion{O}{vi}
emission doublet at 1035~\AA\ using the FUSE satellite. Most recently,
Olofsson et al.~(\cite{Olofsson}) observed extended emission in the
gaseous \ion{Na}{i}\,D lines using high resolution long-slit
spectroscopy.  The observed velocity pattern is consistent with
resonance scattering arising in a disk in Keplerian rotation. The
emission can be traced from less than 30~AU to distances of at least
140~AU. 

This short summary of the debate on $\beta$~Pictoris shows definitely
the need to find suitable gas tracers, which can resolve the question
whether the disks around young A stars are debris disks without any
gas, or protoplanetary disks in which gaseous planet formation may
still proceed. 

This paper concentrates on CO, C and C$^+$ as possible gas tracers in
these disks. Although H$_2$ is contained in our chemical model, we
abstain from a calculation of the near infrared molecular lines, since
it would involve a detailed treatment of the ultraviolet pumping by
stellar and interstellar photons which is beyond the scope of this
paper.  Moreover we do not calculate the Na emission from our
models. Exploratory one-dimensional calculations indicate that in
these disk models the \ion{Na}{i}\,D pumping is not optically thin 
for the stellar radiation. Hence any radiative transfer has to take 
into account not only the stellar radiation field, but also the local 
radiation field arising from the Na atoms within the disk.

In the following sections we shortly summarize the basic properties of
the optically thin disk models and describe the method used to derive
the level populations and the line emission. We present results for a
number of disk models with varying parameters and discuss them in the
light of recent observations. In addition ISO LWS observations of the 
[\ion{C}{ii}] 157.7~$\mu$m line toward Vega and $\beta$~Pictoris are
presented. At the end, suitable tracers to detect the gas in disks 
around A stars are discussed, and line intensities are calculated 
for future instruments and facilities.

\section{The disk models}
\label{diskmodels}
We use here the models for low mass disks surrounding A stars with
stellar spectra similar to $\beta$~Pictoris and Vega presented in an
earlier paper (Kamp \& van Zadelhoff \cite{Kamp2}).  We shortly
summarize the main features of the models and refer to the two
previous papers, Kamp \& Bertoldi (\cite{Kamp1}; hereafter Paper\,{\sc
i}) and Kamp \& van~Zadelhoff (\cite{Kamp2}; hereafter Paper\,{\sc
ii}), for further details.

\subsection{Basic physics of the models}

We assume thin hydrostatic equilibrium models
\begin{equation}
        n(r,z) ~=~n_{\rm i} ~(r/R_{\rm i})^{-2.5}~e^{-z^2/2h^2}
        \label{eq:density}
\end{equation}
with a dimensionless scale-height $H \equiv h/r = 0.15$. The inner and
outer radius of the disk are fixed to $R_{\rm i}=40$~AU and $R_{\rm
o}=500$~AU. The power-law exponent of the disk surface density is $-1.5$,
in reasonable agreement with the literature values of brightness
profiles (Hayashi et al.~\cite{Hayashi}; Dutrey et al.~\cite{Dutrey};
Augereau et al.~\cite{Augereau}) ranging from $-1$ to $-1.5$.

The radiation field $F_\nu$ is composed of a stellar and an
interstellar component. The stellar field is described by an ATLAS9
photospheric model (Kurucz \cite{Kurucz}) for the appropriate stellar
parameters, while the interstellar component is described by a Habing
(\cite{Habing}) field with a flux of $F_{\rm H} = 1.2\times
10^7$~cm$^{-2}$~s$^{-1}$ from 912 \AA \, to 1110 \AA \,penetrating
homogeneously through the entire disk.

The dust temperature follows from radiative equilibrium assuming 
large spherical black body grains of size $a$
\begin{equation}
T_{\rm dust} = 282.5~\left(L_{\ast}/L_\odot\right)^{1/5} 
                             \left(r/{\rm AU}\right)^{-2/5} 
                             \left(a/{\rm\mu m}\right)^{-1/5}~,
\end{equation}
with the stellar luminosity in units of the solar luminosity
$L_\odot$. The assumption of radiative equilibrium is correct
for the optically thin disks described in this paper. For higher mass, 
optically thick disks, the dust-temperature will depend on a detailed 
calculation of the transfer of photons, including scattering, through 
the disk. Scattering will in that case depend strongly on the dust-grain 
size. The use of a grain size distribution would directly affect the
dust temperature and the shielding of UV radiation; the latter is negligible 
for the tenuous disks discussed in this paper. The dust temperature enters 
the chemistry e.g.\ via the formation of H$_2$ and freezing out of CO and
influences the gas temperature in some parts of the disk via IR pumping
of e.g.\ \ion{O}{i} fine structure lines. Moreover the photoelectric
heating of the gas depends on the dust grain size distribution. For the 
purposes of this work, the grain size distribution is approximated 
by an effective dust grain size, which represents the mean properties 
of the dust phase. Hence, in the following a single grain size of 
3~$\mu$m is assumed for all calculations.

The gas temperature and the chemical composition of the circumstellar
disk are derived by coupling the heating and cooling balance and the
equilibrium chemistry. The most relevant heating processes are:
heating by infrared background photons, heating due to the drift
velocity of dust grains, cosmic ray heating, photo-electric heating,
and heating due to formation and photo-dissociation of H$_2$. Cooling,
on the other hand, is dominated by [\ion{O}{i}] and [\ion{C}{ii}]
fine-structure lines as well as the rotational lines of H$_2$ and
CO. The most uncertain process is the drift velocity heating: as
grains are accelerated by the stellar radiation field and moving
through the gas they loose part of their momentum.  The final drift
velocities reached by the dust grains depend strongly on the rotation
of the disk --- this cancels a certain fraction of gravity --- and on
the amount of momentum transfer assumed. In order to bracket reality
we obtained disk models for the two extreme cases: no drift velocity
and a maximum drift velocity
\begin{equation}
v_{\rm drift}^{\rm max} = \left[ \frac{1}{2} \left( \left( 
                      f_{\rm rad}^2 + 
                         v_{\rm gas}^4 \right)^{0.5}
                     - v_{\rm gas}^2 \right) \right]^{0.5}~{\rm cm~s}^{-1}
\label{eq:vdrift}
\end{equation}
where $v_{\rm gas}$ is the thermal velocity of the gas, and $f_{\rm rad}$ is
the radiation pressure on the grains.

The chemical network consists of 47 atomic, ionic, and molecular
species that are related through 260 gas-phase chemical and
photoreactions. A number of reactions is treated in more detail like
H$_2$ and CO photo-dissociation, and C ionization. The only surface
reactions incorporated are H$_2$ formation and freezing out of CO on
cold dust-grain surfaces.  Since we are dealing with large dust
particles, we reduced the H$_2$ formation rate according to the
reduced grain surface area. The abundance of CO ice is due to a
balance between freezing out of gaseous CO and reevaporation of CO
ice. A modified Newton-Raphson algorithm is used to obtain a
stationary solution of the entire chemical network.

\subsection{Grid of A~star disk models}

We chose in Paper\,{\sc i} and {\sc ii} two very prominent
representatives of the class of A stars with disks around them: Vega
and $\beta$~Pictoris. They differ mainly with respect to the strength
of their radiation field, with Vega, spectral type A0V, having an
integrated ultraviolet flux about 6\,000 times larger than the
interstellar UV field at 40~AU, while the $\beta$~Pictoris UV field,
spectral type A5V, is weaker than the interstellar field outwards of
53~AU. At 40~AU it is about 1.8 times the Habing field. These two
radiation fields roughly give the lower and upper limits for A stars.

\begin{table}[ht!]
\caption{\label{modcalculations} Overview of the radiative transfer
calculations. For the models indicated by $\surd$ the lines of C,
C$^+$, and CO were calculated.}
\begin{tabular}{l|ccccc}
\hline
 $F_\nu$          &                        $F_\nu^\star$$^{a}$       & 
 $F_\nu^\star$              & $F_\nu^\star$    & $F_\nu^\star + F_\nu^{\rm IS}$ &
 $F_\nu^\star + F_\nu^{\rm IS}$ \\
 $v_{\rm drift}$  &                                                  &
 $ v_{\rm drift}^{\rm max}$ &  0               & $v_{\rm drift}^{\rm max}$      & 
       0                        \\
 $T_{\rm gas}$    &                        $T_{\rm dust}$            & 
 $\Lambda=\Gamma^{b}$       & $\Lambda=\Gamma$ & $\Lambda=\Gamma$               &
 $\Lambda=\Gamma$               \\
\hline
                  &                                                  &
                            &                  &                                &
                                \\
  Star\,\,         M$_{\rm gas}$        &                            & 
                            &               &                                & 
                                \\
   \hspace*{7mm}   [M$_\oplus$]         &                            &
                            &               &                                &
                                 \\[0.5mm] 
\hline 
 & & & & & \\[-1mm]
A5V\,\,    0.2 & -- & -- & --& $\surd$ & $\surd$ \\ 
A5V\,\,    2.0 & $\surd$ & $\surd$ & $\surd$ & $\surd$ & $\surd$ \\[2mm] 
\hline 
 & & & & & \\[-1mm] 
A0V\,\,    0.2 & --& $\surd$ &$\surd$ & --& --\\
A0V\,\,    2.0 & --& $\surd$ & $\surd$ & -- & --\\[2mm]
\hline
\end{tabular}\\
Note: $^{a}$ $F_\nu^\star$ and $F_\nu^{\rm IS}$ denote the stellar and
interstellar radiation field, respectively \\
\hspace*{9mm}$^{b}$ $\Lambda=\Gamma$ denotes that the
gas temperature is derived from the heating/cooling balance.\\
\end{table}

We use the disk models derived in Paper\,{\sc ii} for both types of
stars and we add two more models for the $\beta$~Pictoris case, where
we include an interstellar UV radiation field penetrating
homogeneously through the disk without any shielding. This is a valid
assumption for disk masses of 2 and 0.2~M$_\oplus$ (see Paper\,{\sc
i}).  Due to the high UV flux of an A0V star the interstellar
radiation field can be neglected for Vega.

The mass of the disk around $\beta$~Pictoris is 54~M$_\oplus$ derived
from the H$_{2}$ S(0) and S(1) emission lines (Thi et al.~\cite{Thi})
and 44~M$_\oplus$ from dust emission, assuming a constant gas-to-dust
mass ratio of 100 (Chini et al.~\cite{Chini:91}).  In this paper only
calculations for lower disk masses are performed, 0.2 and 2.0
~M$_\oplus$, which bracket the observations for $\beta$~Pictoris
(upper limit for the CO emission lines and \ion{C}{i} column densities
from absorption lines). Our models require that the heating and 
cooling processes are optically thin. This prerequisite is
not fulfilled in the $M>2.0~M_\oplus$ disk models since the
[\ion{O}{i}] cooling lines will become optically thick.

\begin{figure*}[ht!] 
\resizebox{\hsize}{!}{\includegraphics{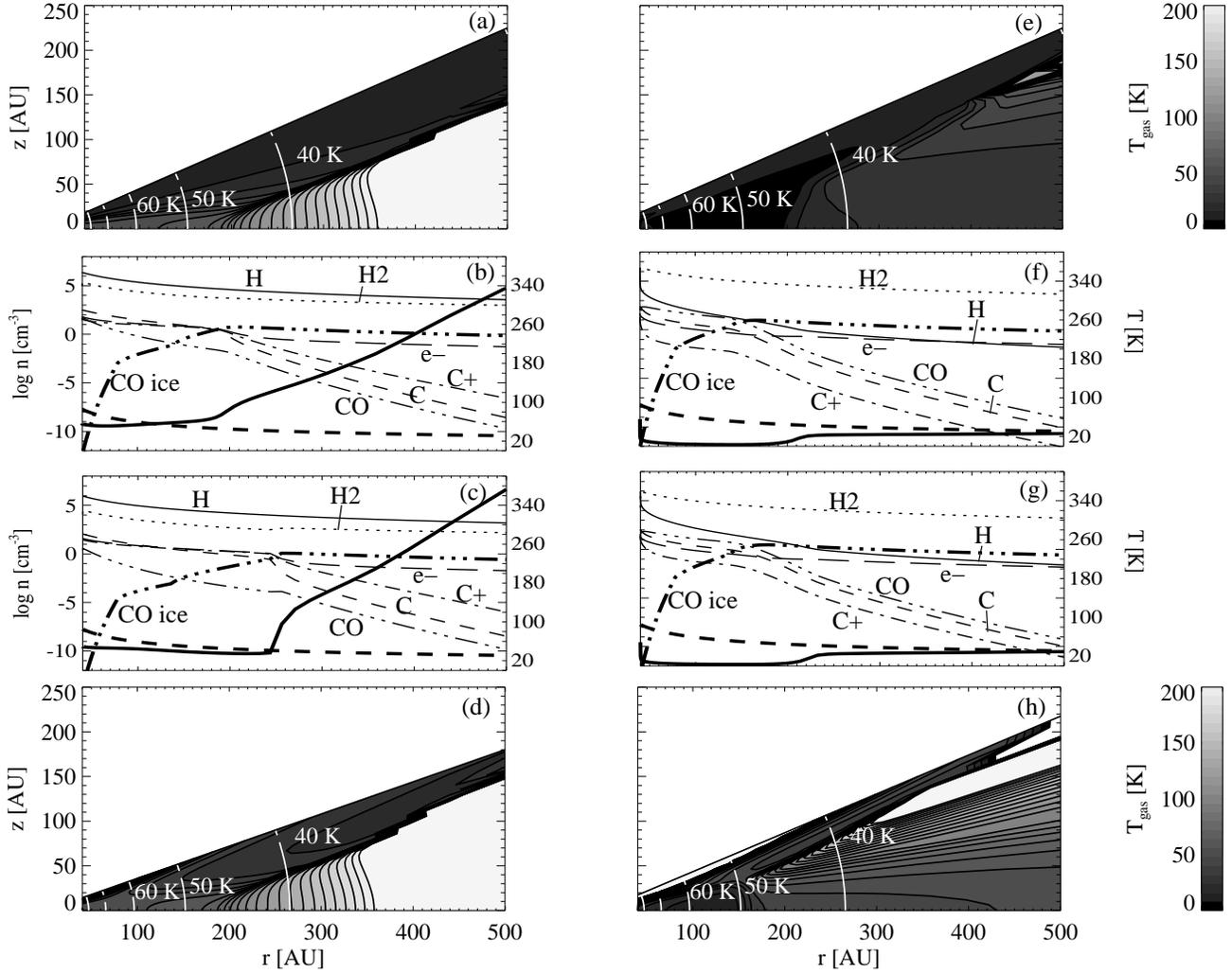}}
\caption{\label{fig: modelsch8} Example of two of the modeled disks
with 2 M$_\oplus$ and stellar UV radiation appropriate for a
$\beta$-Pictoris (A5V) star. The panels (a)-(d) show a disk with
interstellar UV field, the panels (e)-(h) the same disk without the
interstellar UV radiation field. From top to bottom the panels
represent: (a) and (e) the temperature of the gas (greyscale) and the
dust (white contour) for a disk with $v_{\rm drift}=0$; (b) and (f)
the density of the different species at the midplane; (c) and (g) the
density of the different species at 1 scale height ($h=0.15r$); (d)
and (h) temperature of the gas (greyscale) and the dust (white
contour) when $v_{\rm drift}=v_{\rm drift}^{\rm max}$. The adopted
lines in (b), (c), (f) and (g) are: H: solid, H$_{2}$: dotted, C:
dashed, C$^{+}$: dash-dotted, CO: dash-dot-dotted, CO$_{\rm ice}$:
dash-dot-dotted (thick), e-: long dashed, T$_{\rm dust}$: dashed
(thick) and T$_{gas}$: solid (thick).}
\end{figure*}

In order to show the effects of the inclusion of interstellar
radiation as well as to give an overview of the temperature structure
and chemical abundances in the models presented here, two cases are
described in more detail. In Fig. \ref{fig: modelsch8} two disks with
a mass of 2.0~M$_\oplus$ are shown irradiated by a star of spectral
type A5V. The disk model without interstellar UV radiation field
(right hand side of Fig. \ref{fig: modelsch8}) was presented in Paper
II.  The new model includes now also the interstellar UV field (left
hand side of Fig. \ref{fig: modelsch8}). For each model the
temperature structure of the gas and dust are given as well as the
densities of the molecular, atomic and ionic species at the midplane
and at one scale height ($h=0.15r$) of the disk. Both models are
calculated under the assumption that $v_{\rm drift}=0$. To show the
effects of the drift velocity on the gas and dust temperatures, the
corresponding models with $v_{\rm drift}=v_{\rm drift}^{\rm max}$ are
presented as well.  As the chemistry depends only weakly on the
temperature, at least in the temperature range covered in these
models, the densities of each of the species are very similar with or
without taking $v_{\rm drift}$ into account.

In the model without interstellar UV radiation field, hydrogen is
mostly molecular and carbon is in the form of CO. As the
dust temperature drops below 50~K, the critical value assumed in
these models for CO molecules to freeze out onto grain surfaces,
CO ice becomes the dominant carbon bearing species. The abundances
of C and C$^+$ drop according to the freeze out of the gaseous CO.
The inner disk region has gas temperatures around 10~K due to efficient
CO rotational line cooling. Outwards of $\sim 200$~AU, \ion{O}{i}
fine structure line cooling takes over, leading to slightly higher
gas temperatures, $T_{\rm gas} \sim 20$~K. The main heating processes
are pumping of the \ion{O}{i} fine structure levels by infrared
background photons (inner part of the disk) and cosmic ray heating.
If we include drift velocity heating, it becomes the dominant
heating mechanism and as the cooling processes remain the same,
the equilibrium gas temperature is higher. Even though the
chemistry hardly changes due to the higher gas temperature, the
excitation of atoms and molecules depends strongly on T$_{\rm gas}$.
Therefore, detailed modeling of emission lines from circumstellar disks
has to be based on disk models with a realistic determination of the gas 
temperature.

The main difference arising from the inclusion of an interstellar UV
radiation field is the enhanced dissociation of H$_{2}$ and CO and the
enhanced ionisation of C. The CO ice abundance (CO$_{\rm ice}$) is
only marginally affected. In the disk with interstellar UV, CO is
constantly formed and dissociated, but most of the CO freezes
immediately out onto the grains. The CO$_{\rm ice}$ is the sink for
all C-bearing species at larger radii, except for the low density
surface layer, where hardly any CO freezes out. Since the initial
ratio of [O]/[C]$>$ 1, O together with C$^+$ and H$_2$ are the most
important coolants available. Due to the enhanced UV flux, H$_2$
dissociation and formation are the most important heating sources,
leading to higher equilibrium gas temperatures than in the previous
case without the interstellar UV radiation field. As soon as C$^+$ is lost
as a coolant because of its incorporation into CO ice, the gas
temperatures rise quickly to values in excess of 100~K. Incorporating
the drift velocity into this model does not alter the gas temperature,
because H$_2$ dissociation and formation heating are more efficient
than drift velocity heating.

A complete list of all adopted models and performed radiative
transfer calculations can be found in Table~\ref{modcalculations}.

\section{Radiative transfer calculations}

For the line radiative transfer the Monte Carlo code developed by
Hogerheijde \& van der Tak (\cite{HvdT00}) is used. The disk models
described in Sect.~\ref{diskmodels} are in\-terpolated on a
cylindrical grid ($26\times 10$ grid cells), with a logarithmic radial 
grid in order to smoothly follow the density and temperature gradients. 
The size of the grid is small due to computational 
constraints. A comparison run with twice the number of grid cells gave a
less than 20\% difference in integrated line-intensity, comparable to the 
observational errors. The calculations concentrate on the non 
local-thermal-equilibrium (NLTE) intensities of
the fine structure lines of C and C$^+$ and the rotational transitions
of CO.  Table~\ref{lines} shows the respective line data and
beam sizes used for convolution in the section on comparison to
observations (Sec. \ref{sec: comparch8}). The following paragraphs
explain the calculations for the individual species in more detail.

\begin{table}[ht!]
\caption{\label{lines} Adopted line data for the fine-structure lines
         of C and C$^+$, and the rotational lines of CO.}
\begin{tabular}{llcll}
\hline   & Line &          & $A_{\rm ul}$ & Beam \\
         &      &          & [s$^{-1}$]   & [\arcsec] \\ 
\hline\\ 
C     & 809.3 GHz     & $^{3}$P$_{2}$-$^{3}$P$_{1}$     & $2.65\,10^{-7}$ &  6$^{a}$;8$^{b}$;\\
      &               &                                 &                 & 26$^{c}$ \\ 
      & 492.2 GHz     & $^{3}$P$_{1}$-$^{3}$P$_{0}$     & $7.93\,10^{-8}$ & 10$^{a}$;13$^{b}$;\\
      &               &                                 &                 & 44$^{c}$\\[2mm] 
C$^+$ & 157.74 $\mu$m & $^{2}$P$_{3/2}$-$^{2}$P$_{1/2}$ & $2.29\,10^{-6}$ & 11$^{c}$;16$^{d}$\\[2mm] 
CO    & 115.3 GHz     & 1-0                             & $7.17\,10^{-8}$ & 43$^{a}$\\ 
      & 230.5 GHz     & 2-1                             & $6.87\,10^{-7}$ & 22$^{a}$\\
      & 345.8 GHz     & 3-2                             & $2.48\,10^{-6}$ & 14$^{a}$\\
\hline
\end{tabular}\\
$^{a}$15 m telescope (SEST, JCMT),$^{b}$12 m telescope (APEX),
$^{c}$HIFI (Herschel), $^{d}$SOFIA \\
\end{table}

In the above mentioned code, the equations of statistical equilibrium
are solved in an iterative fashion, where all photons start at the
outer boundary with an intensity given by the 2.728~K cosmic
background radiation. The abundances of the trace species under
consideration as well as their most important collision partners are
taken from our stationary disk models (Paper\,{\sc ii} and
Fig. \ref{fig: modelsch8}).  

The code allows the use of two different collision partners and
Table~\ref{tabcoll} shows the most important partners for each species
and the respective references for the collisional rate
coefficients. Stimulated absorption and emission by far-infrared
radiation from dust is not significant for these species. For C--H$_2$
and C$^+$--H$_2$ collisions, we assume that the ortho-to-para ratio of
molecular hydrogen is determined by the local gas temperature.  In the
case of C$^+$, scaled H$_2$ collision cross sections for atomic
hydrogen are used; hence the total density of the collision partner is
given by the weighted sum of the H$_2$ and H density, namely $n({\rm
H_2})+n({\rm H})/0.57$, and $n(e)$ is considered as the second
partner. Comparing CO--H and CO--e$^-$ collision cross sections, it is
found that for $n({\rm H})/n({\rm e^-})<10^3$, collisions with
electrons are more important than collisions with neutral hydrogen. In
the case of all A0V models and the A5V models with interstellar
radiation, $n$(H) is the main collision partner for CO; in the
remaining models not enough H$_{2}$ is dissociated and collisions with
electrons dominate over atomic hydrogen collisions.

\begin{table}
\caption{\label{tabcoll} Collision partners for the different species and the
 respective references for the collisional rate coefficients.}
\begin{tabular}{lll}
\hline
Species  & Collision partner & Reference  \\
\hline\\
 C     & H$_2$               & Schr\"{o}der et al.~(\cite{Schroeder}) \\
       & H                   & Launay \& Roueff (\cite{Launay})       \\ 
 C$^+$ & H$_2$               & Flower (\cite{Flower})           \\
       & H                   & Flower (\cite{Flower})$^{a}$             \\
       & e$^-$               & Mendoza (\cite{Mendoza}),               \\
       &                     & Keenan et al.~(\cite{Keenan})          \\ 
 CO    & H$_2$               & Schinke et al.~(\cite{Schinke})      \\
       & H                   & Chu \& Dalgarno (\cite{Chu})\\
       & e$^-$               &      \\
\hline
\end{tabular}\\
$^{a}$Scaled from H$_{2}$ collision rates. See text for details.\\
\end{table}

The resulting populations at each position in the disk are used to
compute the sub-millimeter line profiles of CO, C, and C$^+$ using a
ray tracing program which calculates the sky brightness
distribution. Observations of circumstellar ultraviolet absorption
lines of CO and C around $\beta$~Pictoris reveal a line-broadening
parameter $b=1.3$~km~s$^{-1}$ (Roberge et al.~\cite{Roberge}).  This
value is used for all model calculations.

\section{Results}
\label{results}
As a first result the line profiles are calculated for a beam with the
same size as the apparent diameter of the disk on the sky. The beam is
centered on the stars and has a size of 50.5\,\arcsec\ for simulating
$\beta$~Pictoris and 128.9\,\arcsec\ for Vega at their appropriate
distances, 19.82 and 7.76 parsec, respectively. The adopted inclinations
are taken from $\beta$~Pictoris (edge on) and Vega (face on).

Since the emission lines are optically thin, the integrated
intensities can be converted directly to any inclination. These
results can therefore be used for any disk around an A star as long as
the total integrated flux $\int T_{\rm mb}dv \times
\Omega_{\rm mb}$ is kept constant.  In the following the presentation of
the results is separated into two sections. The first deals with 
A5V and the second with A0V stars.

\subsection{Spectral type A5V}

\begin{figure}[ht!]
\resizebox{\hsize}{!}{\includegraphics{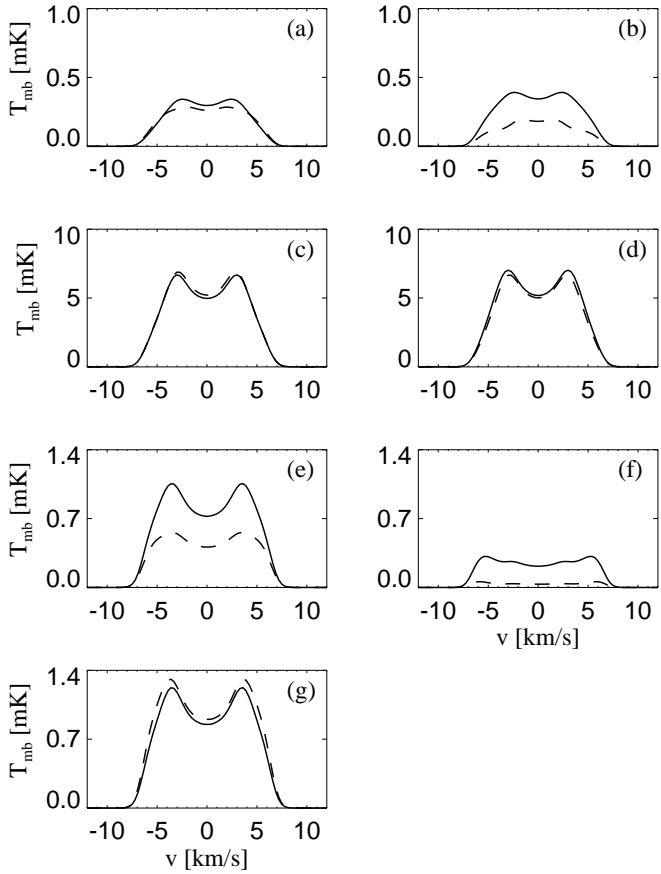}}
\caption{\label{bpic_C_overview}[\ion{C}{i}] fine-structure line
         profiles (main beam brightness temperature versus rest velocity) in
         different A5V models: (a) 0.2~M$_\oplus$ model with $v_{\rm
         drift}=v^{\rm max}_{\rm drift}$ and an interstellar radiation
         field, (b) 0.2~M$_\oplus$ model with $v_{\rm drift}=0$ and
         interstellar radiation field, (c) 2~M$_\oplus$ model with
         $v_{\rm drift}=v^{\rm max}_{\rm drift}$ and interstellar
         radiation field, (d) 2~M$_\oplus$ model with $v_{\rm
         drift}=0$ and interstellar radiation field, (e) 2~M$_\oplus$
         model with $v_{\rm drift}=v^{\rm max}_{\rm drift}$, (f)
         2~M$_\oplus$ model with $v_{\rm drift}=0$, (g) 2~M$_\oplus$
         model with $T_{\rm gas}=T_{\rm dust}$. The solid line denotes
         the $^{3}$P$_{1}$--$^{3}$P$_{0}$ transition at 492.2~GHz (609.13~$\mu$m)
         and the dashed line the $^{3}$P$_{2}$--$^{3}$P$_{1}$
         transition at 809.3~GHz (370.42~$\mu$m.)}
\end{figure}

\begin{figure}[ht!]
\resizebox{\hsize}{!}{\includegraphics{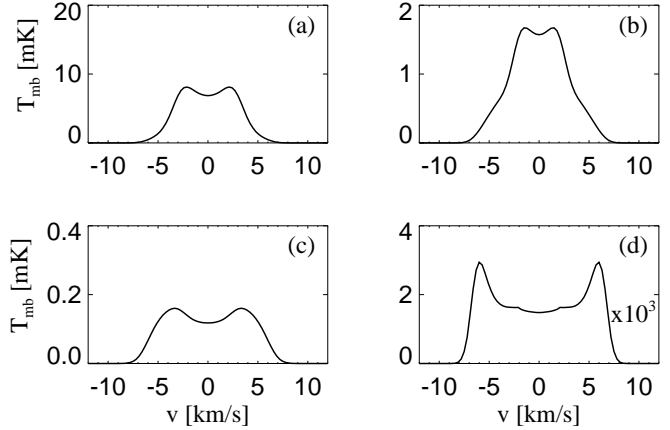}}
\caption{[\ion{C}{ii}]\,157.74~$\mu$m fine-structure line
         profiles (main beam brightness temperature versus rest velocity) in
         different A5V star models: (a) 2~M$_\oplus$ model with
         $v_{\rm drift}=v^{\rm max}_{\rm drift}$ and interstellar
         radiation field, (b) 0.2~M$_\oplus$ model with $v_{\rm
         drift}=v_{\rm drift}^{\rm max}$ and interstellar radiation
         field, (c) 2~M$_\oplus$ model with $v_{\rm drift}=v_{\rm
         drift}^{\rm max}$, and (d) 2~M$_\oplus$ model with $v_{\rm
         drift}=0$ (note the flux is multiplied by 10$^{3}$).}
\label{bpic_C+_lines}
\end{figure}

 We choose atomic carbon as an example to illustrate the effects of the various
model parameters on the resulting line profiles
(Fig.~\ref{bpic_C_overview}). The shapes of the emission-line profiles
show a double peaked structure, with subtle differences due to
excitation and/or abundances (e.g., Figs.~\ref{bpic_C_overview}f and
\ref{bpic_C+_lines}b). 

For the 2M$_\oplus$ model including the interstellar UV radiation
field (Fig.~\ref{bpic_C_overview}c,d), the gas temperature rises above
50~K resulting in a
$^{3}$P$_{2}$--$^{3}$P$_{1}$/$^{3}$P$_{1}$--$^{3}$P$_{0}$ ratio of
approximately unity. The carbon density rises by an order of
magnitude compared to the other 2M$_\oplus$ models, because CO is no
longer the dominant reservoir of carbon due to the additional
photodissociation. There is no difference in gas temperature between
the calculation with or without drift velocity heating, as the
temperature balance is in this case entirely determined by H$_2$
formation and dissociation (Fig.~\ref{fig: modelsch8}).

For the cases with only the stellar radiation field, the difference
between the 2~M$_\oplus$ models with (Fig.~\ref{bpic_C_overview}e) and
without (Fig.~\ref{bpic_C_overview}f) drift velocity is due to the gas
temperature. $T_{\rm gas}$ drops below 20~K (Fig.~\ref{fig:
modelsch8}) for a large part of the disk without any drift velocity
heating, keeping most of the carbon in the ground state. Hence the
lines are significantly weaker than with drift velocity heating
included.

In the case where $T_{\rm gas}=T_{\rm dust}$
(Fig.~\ref{bpic_C_overview}g), the $^{3}$P$_{2}$--$^{3}$P$_{1}$ line
is stronger than the $^{3}$P$_{1}$--$^{3}$P$_{0}$ line contrary to the
other 2~M$_\oplus$ models (Figs.~\ref{bpic_C_overview}e, f) which include
heating and cooling. This is a direct result of the
temperature. In the models with heating and cooling, the gas
temperature is well below 30~K in the inner 150~AU, while it stays
above 50~K if $T_{\rm gas}=T_{\rm dust}$ is assumed. The $^3P_2$
fine-structure level of C lies at 62 K and is only significantly
populated at temperatures above 30~K.

If the disk mass is reduced by a factor of 10~
(Figs.~\ref{bpic_C_overview}a, b), the carbon density decreases by a
slightly larger factor, because the ionization balance shifts towards
C$^+$ as shielding decreases in lower mass models.  The temperature
stays above 40~K in the inner 100~AU of the disk, giving rise to
equally strong emission lines.  This is not the case if the
drift-velocity heating is absent. In that case the temperature
drops below 30~K causing the $^{3}$P$_{2}$--$^{3}$P$_{1}$ line to
become much weaker and the $^{3}$P$_{1}$--$^{3}$P$_{0}$ slightly
higher.

\begin{figure}[ht!]
\resizebox{\hsize}{!}{\includegraphics{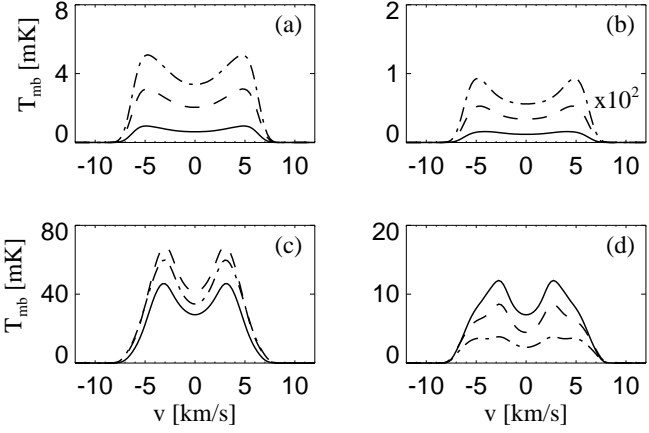}}
\caption{CO rotational line profiles (main beam brightness 
  temperature versus rest velocity) in different A5V star models: (a)
  2~M$_\oplus$ model with $v_{\rm drift}=v_{\rm drift}^{\rm max}$ and
  interstellar radiation field, (b) 0.2~M$_\oplus$ model with $v_{\rm
  drift}=v_{\rm drift}^{\rm max}$ and interstellar radiation field
  (note the flux is multiplied by 10$^{2}$), (c) 2~M$_\oplus$ model
  with $v_{\rm drift}=v_{\rm drift}^{\rm max}$, and (d) 2~M$_\oplus$
  model with $v_{\rm drift}=0$. Solid line: $J$=1--0, dashed line:
  $J$=2-1, dot-dashed line: $J$=3--2.}
\label{bpic_CO_lines}
\end{figure}

For C$^+$ and CO we concentrate on four models, the 2 and 0.2~
M$_\oplus$ model with interstellar UV radiation and $v_{\rm
drift}=v_{\rm drift}^{\rm max}$, and the 2~M$_\oplus$ models with
$v_{\rm drift}=v_{\rm drift}^{\rm max}$ and $v_{\rm drift}=0$ without
interstellar radiation. The combination of these models shows all the
possible influences in the models. Comparisons between panels (a) and
(c) show the influence of the interstellar UV radiation field, between
(a) and (b) of the disk mass, and between (c) and (d) of the
drift-velocity heating on the emission lines of C$^+$ and CO.

The influence of the interstellar radiation field on the temperature
and chemical abundances can be seen in Fig.~\ref{fig: modelsch8}. The
CO abundance is reduced due to enhanced photodissociation, and the
C$^+$ line therefore becomes much stronger. Moreover, the lower CO
transitions $J$=1--0 and $J$=2--1 are generally weaker than the
$J$=3--2 line. This is due to typical gas temperatures in excess of
30~K (the $J$=3 level lies at 33~K) due to the interstellar radiation
field. The level populations were calculated using the full
statistical equilibrium calculation. This has only a marginal effect
on the three lowest levels of CO for the 2.0 M$_\oplus$ model.  Only
in the surface layers of the disk are the densities low enough to
prevent collisionally excitation of $J$=3. These slight differences
are therefore only marginally visible in the line-profiles and not
distinguishable with current telescope facilities.

The influence of the drift-velocity on the line emission is large in
the case without interstellar UV radiation. Due to the extremely low
gas-temperatures in the inner regions of the disk, where the CO
density is highest, the $J$=1 level is excited more efficiently than
the higher levels, making the lowest transition the strongest
(Fig.~\ref{bpic_CO_lines}d).

In the case of the lower mass (0.2 M$_\oplus$) disk, the level
populations do show clear NLTE effects on the level populations, even
close to the midplane. The total emission, however, is extremely
small. Whereas the C and C$^{+}$ line emission scale roughly with
mass, the CO line emission is reduced by nearly three orders of
magnitude due to enhanced dissociation in the lower mass disk.

\subsection{Spectral type A0V}

Since we adopted the disks around A0V stars to be seen face on, no
coherent velocities are seen in the line of sight and only the
micro-turbulent ($b=0.91$~km~s$^{-1}$) and thermal velocities
remain.

\begin{figure}[ht!]
\resizebox{\hsize}{!}{\includegraphics{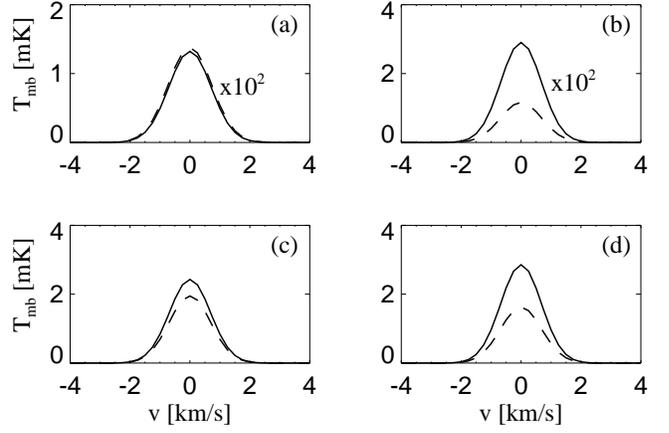}}
\caption{[\ion{C}{i}] fine-structure line profiles (main beam brightness 
  temperature versus rest velocity) in different A0V models:
  (a) 0.2~M$_\oplus$ model with $v_{\rm drift}=v_{\rm drift}^{\rm
  max}$, (b) 0.2~M$_\oplus$ model with $v_{\rm drift}=0$, (c)
  2~M$_\oplus$ model with $v_{\rm drift}=v_{\rm drift}^{\rm max}$, (d)
  2~M$_\oplus$ model with $v_{\rm drift}=0$. The solid line denotes
  the $^{3}$P$_{1}$--$^{3}$P$_{0}$ transition at 609.13~$\mu$m and the
  dashed line the $^{3}$P$_{2}$--$^{3}$P$_{1}$ transition at
  370.42~$\mu$m. Note that the fluxes in panels (a) and (b) are
  multiplied by 10$^{2}$.}
\label{vega_C_overview}
\end{figure}

All line profiles show a single gaussian emission peak, and
are generally orders of magnitude weaker for C and CO than in the A5V
models. Due to the high UV flux from the A0V stars, most of the CO in
the gas phase is dissociated. The atomic carbon is subsequently ionized enhancing
the C$^{+}$ abundance and its emission by a few orders of magnitude.

Similar to the A5V stars, we illustrate the differences arising from
the various models by using C as an example. The models with
drift-velocity heating are generally warmer than those without. This
is nicely illustrated in Fig.~\ref{vega_C_overview}, where the
$^{3}$P$_{2}$--$^{3}$P$_{1}$/$^{3}$P$_{1}$--$^{3}$P$_{0}$ line ratio
is higher for the models with $v_{\rm drift}=v_{\rm drift}^{\rm
max}$. The temperature effect is more pronounced in the 0.2~M$_\oplus$
model, as the ratio is larger than unity. For the $v_{\rm drift}=0$
models, the cooling is more efficient in the low-mass disk giving a
relatively large $^{3}$P$_{1}$--$^{3}$P$_{0}$ line emission compared
to the higher transition. In the low mass models, the ionization
fraction is much higher due to a general lack of shielding. This leads
to C emission that is a factor of 100 lower than in the 2~M$_\oplus$
models.

\begin{figure}[ht!]
\resizebox{\hsize}{!}{\includegraphics{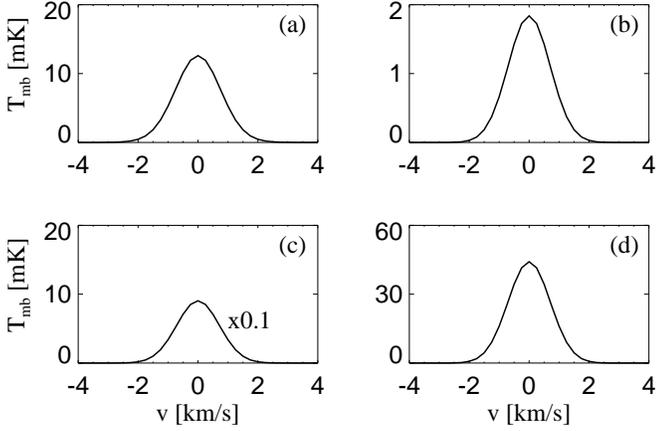}}
\caption{[\ion{C}{ii}]\,157.74~$\mu$m fine-structure line profiles 
         (main beam brightness temperature versus rest velocity) in
         different A0V star models: (a) 0.2~M$_\oplus$ model with
         $v_{\rm drift}=v_{\rm drift}^{\rm max}$, (b) 0.2~M$_\oplus$
         model with $v_{\rm drift}=0$, (c) 2~M$_\oplus$ model with
         $v_{\rm drift}=v_{\rm drift}^{\rm max}$ (note that the flux
         is multiplied by 0.1), and (d) 2~M$_\oplus$ model with
         $v_{\rm drift}=0$.}
\label{vega_C+_overview}
\end{figure}

For C$^+$ the differences are entirely due to the excitation. The
C$^+$ densities scale with the mass, but the population of the upper
level ($91.2$~K) depends strongly on the gas temperature. For CO, the
$J$=3--2 line is the strongest rotational transition in the models
with $v_{\rm drift}=v_{\rm drift}^{\rm max}$ and roughly equal to the
2--1 line in the other two models.

\begin{figure}[ht!]
\resizebox{\hsize}{!}{\includegraphics{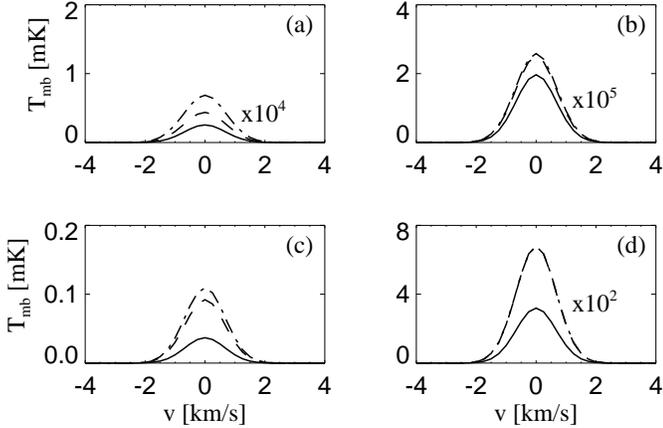}}
\caption{CO rotational line profiles (main beam brightness 
  temperature versus rest velocity) in different A0V star models: (a)
  0.2~M$_\oplus$ model with $v_{\rm drift}=v_{\rm drift}^{\rm max}$,
  (b) 0.2~M$_\oplus$ model with $v_{\rm drift}=0$, (c) 2~M$_\oplus$
  model with $v_{\rm drift}=v_{\rm drift}^{\rm max}$, and (d)
  2~M$_\oplus$ model with $v_{\rm drift}=0$.  The solid line is
  $J$=1--0, dashed line $J$=2-1, dot-dashed line $J$=3--2. Note that
  the fluxes are multiplied by the factors indicated.}
\label{vega_CO_overview}
\end{figure}

\subsection{Integrated intensities}

The line profiles presented in the above sections are calculated for a
single position using different beam-sizes. Therefore they do not
reveal in which part of the disk the emission arises, nor at which
position the beam should be centered to pick up the maximum emission
from the disk.  To illustrate the radial dependence of the emission,
the disk is scanned with a small beam over the disk midplane,
2\,\arcsec\ in the case of the edge-on A5\,V star and 5\,\arcsec\ in the
case of the pole-on A0\,V star.

\begin{figure}[ht!]
\resizebox{\hsize}{!}{\includegraphics{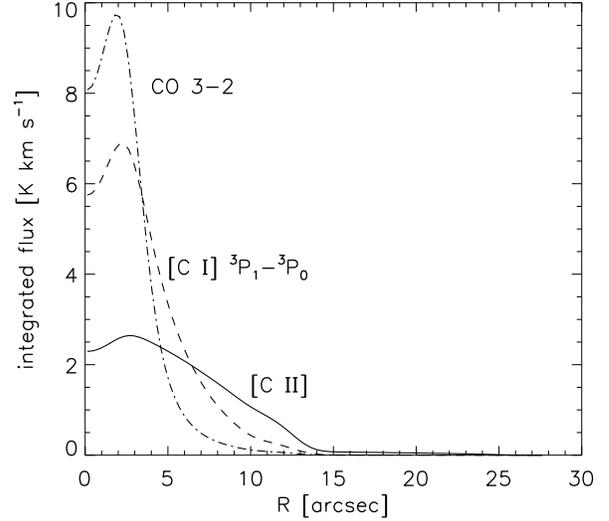}}
\caption{Integrated flux along the disk midplane of the 2~M$_\oplus$
         model illuminated by an A5\,V star, with $v_{\rm drift}=0$
         and interstellar radiation field using a 2\,\arcsec\ beam size:
         CO 3--2 (solid line), [\ion{C}{i}]
         $^{3}$P$_{1}$--$^{3}$P$_{0}$ (dashed line), and [\ion{C}{II}]
         (157.74 $\mu$m, dash-dotted line). The distance to the star
         is taken to be 19.8 pc.}
\label{bm2.0_hclp_IS_integrated}
\end{figure}

Fig.~\ref{bm2.0_hclp_IS_integrated} shows that the maximum emission
for the A5V star for all three tracers, CO, C and C$^+$ is at
2.5\,\arcsec, corresponding to 50~AU at a distance of 19.8~pc. In the
inner disk the particle densities are highest whereas outwards the
density drops with $r^{-2.5}$. Beyond 10\,\arcsec, i.e. 200~AU, carbon
is mainly in the form of CO ice (see Fig.~\ref{fig: modelsch8}) and
hence the emission gradually drops off. 

In the case of the A0\,V star the stellar UV flux is strong enough to
keep C$^{+}$ as the main emitting carbon-species throughout the
disk. The emission is strongest between 5 and 10\,\arcsec\ (40-80 AU)
for C$^{+}$ and C and between 10-15\,\arcsec\ (80-120) for CO. Closer
to the star the beam is diluted due to the inner hole of 40 AU, hence
the rapid decline in emission close to the star.

\begin{figure}[ht!]
\resizebox{\hsize}{!}{\includegraphics{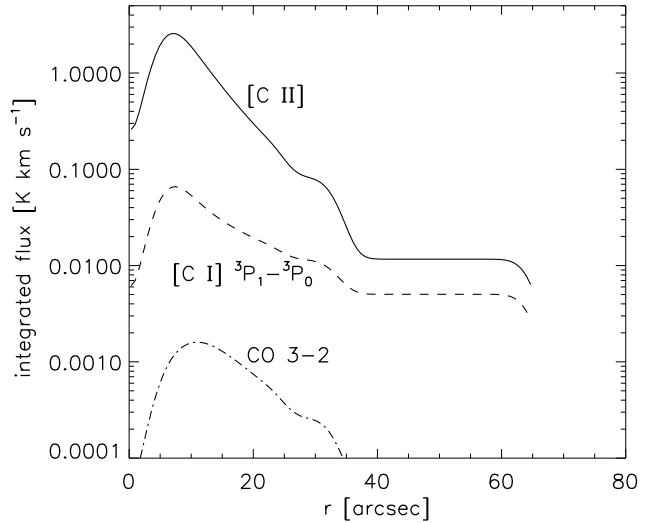}}
\caption{Integrated flux along the disk radius of the A0\,V star,
         2~M$_\oplus$ model with $v_{\rm drift}=0$ with a 5\,\arcsec\
         beam size: CO 3--2 (solid line), [\ion{C}{i}]
         $^{3}$P$_{1}$--$^{3}$P$_{0}$ (dashed line), and [\ion{C}{ii}]
         (157.74 $\mu$m dash-dotted line). The distance to the star is
         taken to be 7.8 pc.}
\label{vm2.0_hclp_integrated}
\end{figure}

\section{[\ion{C}{ii}] observations of Vega and $\beta$~Pictoris}
\label{cii_obs}

The Infrared Space Observatoty (ISO) data archive was used to
retrieve all observations made with the Long Wavelength Spetrometer
(LWS) in a radius of 15$'$ toward Vega and $\beta$~Pictoris. This 
yielded data from projects by MBARLOW and RSTARK taken in full 
grating scan mode (AOT L01) and grating range scan mode (LWS AOT 
L02). The spectral sampling interval of all data from these projects is 
1/4 of a resolution element. The number of grating scans (forward and 
backward) varied between 6 and 24. Each scan yields simultaneous data
accross ten detectors. The total integration time varied between 1172~s and 
3592~s per AOT. We only considered the data from detector LW04 whose
central wavelength is at $\sim 160 \mu$m (grating in nominal position). 
The effective beam size of this detector is 78$''$ (FWHM), the spectral 
resolution is about 0.6~$\mu$m which corresponds to a velocity 
resolution of about 1150~km~s$^{-1}$ at 157.7~$\mu$m (ISO Handbook 
Vol. IV; Version 1.2).

The data were automatically processed through the LWS pipeline (OLP version 10), 
yielding LWS Auto analysis (LSAN) files which contain the flux and wavelength
calibrated spectrum of an AOT. The absolute flux calibration is estimated at 
about 30\% (Swinyard et al. \cite{Swinyard}). The wavelength accuracy for the LW detectors 
is $\sim 0.15~\mu$m (ISO Handbook Vol. IV; Version 1.2). Final data reduction 
was done manually using the ISO spectral Analysis Package (ISAP) Version 2.1.
Glitches and their related decays due to Cosmic Rays were removed from
each individual grating scan. A medium clip was subsequenly applied to
all scans for all data in a bin which is more than 2.5\,$\sigma$ larger 
or smaller than the median. Since the observed sources are not bright nor
larger than the LWS beam, no defringing was applied. Subsequently an average 
to the mean of all subscans was done for each bin. A linear baseline was fitted 
to the averaged spectrum and the line fluxes were estimated through fitting
of a gaussian profile with a FWHM comparable to the instrumental resolution. 

\begin{figure}[ht!]
\resizebox{\hsize}{!}{\includegraphics{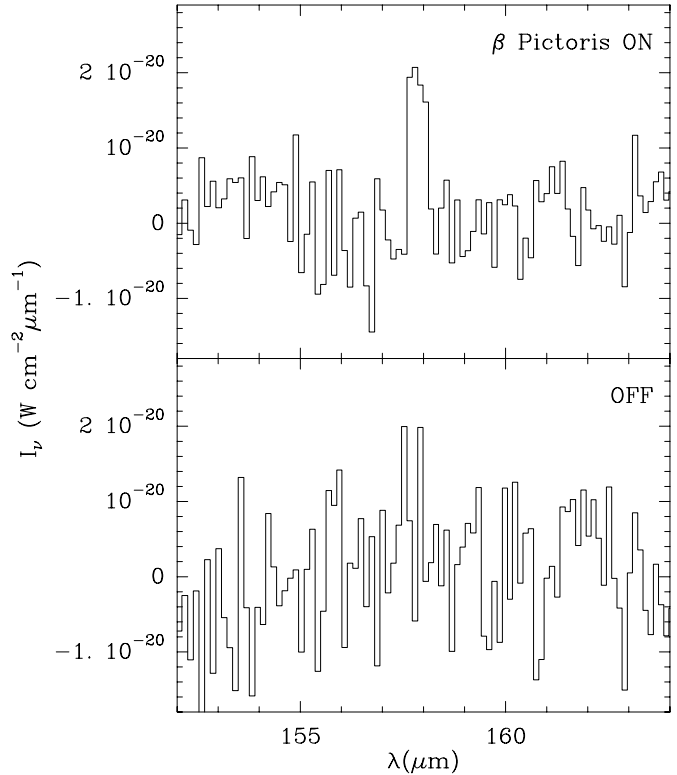}}
\caption{ISO LWS spectra of the [\ion{C}{ii}] 157.7~$\mu$m line for
         $\beta$~Pictoris (ON 05h47m17.11s -51d03'59.5'' and OFF
         05h47m22.66s -50d57'28.5'', J2000).}
\label{fig:cii_obs}
\end{figure}

The resulting spectra are presented in Fig.~\ref{fig:cii_obs}. The spectrum 
toward Vega does 
not reveal any spectral line, a 2\,$\sigma$ upper limit to the [\ion{C}{ii}] 
line intensity was estimated at $<1.6\times 10^{-20}$~W~cm$^{-2}$~$\mu$m$^{-1}$ 
in a 0.15 $\mu$m bin. The line-of-sight toward $\beta$~Pictoris shows a 
$4\,\sigma$ feature with a maximum intensity of $1.8 \times 10^{-20}$~W~cm$^{-2}$~$\mu$m$^{-1}$ 
and an integrated intensity of about $1 \times 10^{-20}$~W~cm$^{-2}$ centred at 
157.85~$\mu$m, which agrees with the wavelength of the [\ion{C}{ii}] line within the LWS 
accuracy. The off-spectrum does not reveal any emission at this frequency with a
$2\,\sigma$ upper limit of $1.0 \times 10^{-20}$~W~cm$^{-2}$~$\mu$m$^{-1}$ 
in a 0.25~$\mu$m bin which indicates that there is no contamination by an 
extended interstellar [\ion{C}{ii}] emission component. However, a contribution of
weak interstellar [\ion{C}{ii}] cannot be fully excluded since the integration time
used for the OFF position is a factor 2.5 lower than used toward $\beta$~Pictoris.

\section{Comparison of models with observations}
\label{sec: comparch8}

\begin{figure}[ht!]
\resizebox{\hsize}{!}{\includegraphics{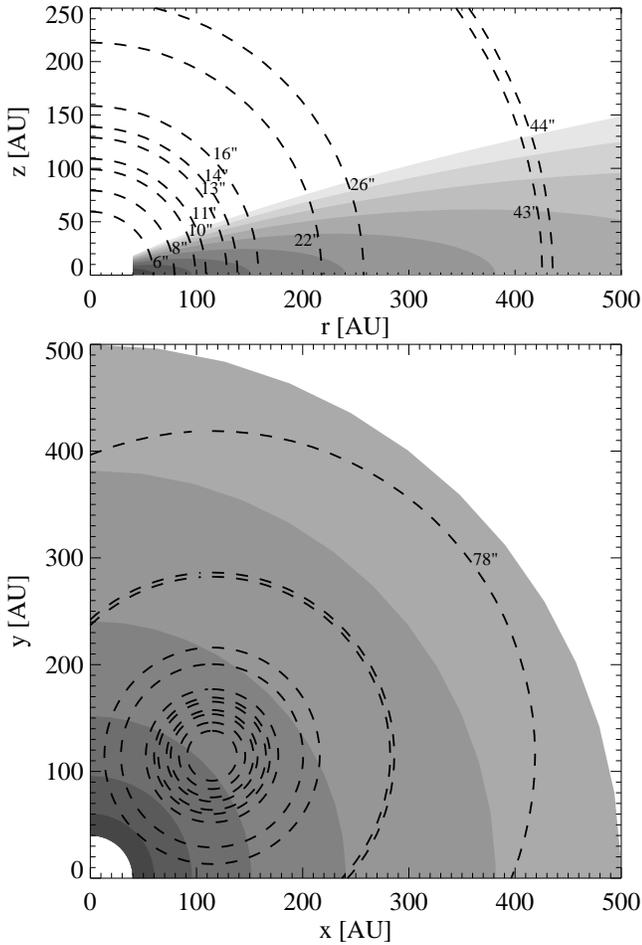}}
\caption{Density distributions for $\beta$~Pictoris (top) and Vega
(bottom). The telescope beams used in the calculations are
superposed as dashed circles. See Table \ref{lines} for the corresponding
telescopes. The beam sizes in the lower panel correspond to those in
the top panel.}
\label{beam}
\end{figure}

The attempts to detect the gas component in low mass disks around A
stars concentrate mostly on prominent A stars like $\beta$~Pictoris
and Vega. In order to compare the models to the observations, we have
chosen the same beam sizes as used in the different observations.
Table~\ref{lines} shows the beam sizes for each line and telescope,
while Fig.~\ref{beam} illustrates the areas of the disk contained in
the different beams for $\beta$~Pictoris (edge-on at a distance of
19.8~pc) and Vega (pole-on at a distance of 7.8~pc). For Vega a radial
offset of 14.7\,\arcsec\ (114~AU) is applied because the smallest beams
contain only the star and the inner hole of the disk model. Observations
by Holland et al. (\cite{Holland:1998}) and subsequent modeling by
Dent et al. (\cite{Dent:00}) substantiate the presence of these
holes.

\subsection{\boldmath{$\beta$}-Pictoris}

\begin{figure}[ht!]
\resizebox{\hsize}{!}{\includegraphics{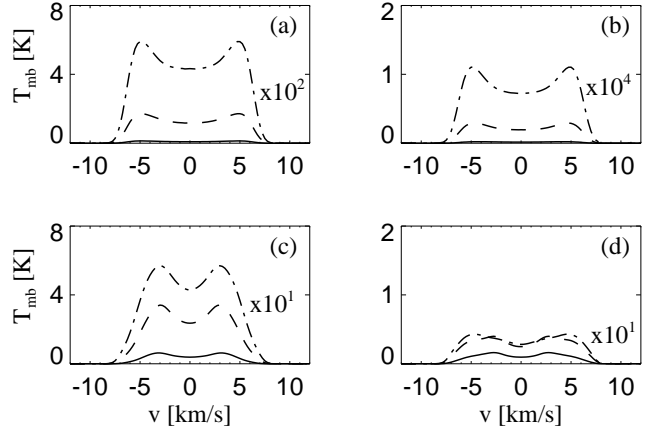}}
\caption{CO rotational line profiles (main beam brightness 
  temperature versus rest velocity) in different $\beta$~Pictoris
  models convolved with the observational beams: (a) 2~M$_\oplus$
  model with $v_{\rm drift}=v_{\rm drift}^{\rm max}$ and interstellar
  radiation field, (b) 0.2~M$_\oplus$ model with $v_{\rm drift}=v_{\rm
  drift}^{\rm max}$ and interstellar radiation field, (c) 2~M$_\oplus$
  model with $v_{\rm drift}=v_{\rm drift}^{\rm max}$ without
  interstellar radiation field, and (d) 2~M$_\oplus$ model with
  $v_{\rm drift}=0$. The solid line is $J$=1--0 (43\,\arcsec ), the
  dashed line $J$=2--1 (21\,\arcsec ), and the dot-dashed line $J$=3--2
  (14\,\arcsec ). Note that the fluxes are multiplied by the factors
  indicated.}
\label{co_beta_pic}
\end{figure}

For $\beta$~Pictoris, the CO $J$=1--0 and 2--1 lines have been
searched several times with the SEST (Savoldini \& Galletta
\cite{Savoldini}; Liseau \& Artymowicz \cite{Liseau1}) and the JCMT
(Dent et al.~\ \cite{Dent}).  In all cases only an upper limit on the
CO column density was derived, the most strict being $N_{\rm CO} <
3\,10^{14}$~cm$^{-2}$ so far. This is deduced from the CO 2--1 upper
limit found by Liseau \& Artymowicz (\cite{Liseau1}), where they
obtain a noise level of $T_{\rm rms}=11$~mK using a binning of
0.9~km~s$^{-1}$.  With a binning of 5~km~s$^{-1}$, which corresponds
to the halfwidth of the CO line, the noise level goes down to
4~mK. To compare these limits to the models, T$_{\rm rms}$ has to
be converted using the main beam efficiency $\eta_{\rm mb}=0.38$ of the 
SEST at the corresponding wavelength. The models presented show that 
the $J$=2--1 transition in the 
$\beta$~Pictoris model with 2~M$_\oplus$ and the interstellar radiation 
field reaches a flux of 12.2~mK with a binning of 5~km~s$^{-1}$, 
while the $J$=1--0 line reaches only 0.9~mK (Fig.~\ref{co_beta_pic}a).  
At the typical disk temperatures 
of 50~K in the models, mostly the $J$=2 and $J$=3 levels are populated. 
The predicted $J$=2--1 line would not have been detected 
by the latest SEST observations, which have $T_{\rm mb,rms}=T_{\rm rms}/\eta_{\rm mb}=10.5$~mK.

The models show that in fact the most favorable line to observe is the
$J$=3--2 line. A spectrum for this line, taken from the JCMT public
archive, has a noise level of $T_{\rm rms}=27$~mK after smoothing to
1.1~km~s$^{-1}$ velocity bins. A much deeper integration should be
possible with the dual polarization B3 receiver. The current limit is
a factor of two lower than the modeled line (Fig.~\ref{co_beta_pic}a,
dot-dashed line).

The 2~M$_\oplus$ disk models have significantly lower mass than the
value of $\sim50 M_\oplus$ inferred from the dust using a gas/dust
ratio of 100:1. Scaling the mass up by that factor ($\times 25$), both
the $J$=2--1 and $J$=3--2 line should have been easily detected
even though the CO gas abundance is unlikely to scale linearly due to
enhanced freeze-out.

Comparison between Figs.~\ref{co_beta_pic}c and d reveals that without
drift velocity heating the disk becomes much cooler and the lines
weaker. Even the line ratios are affected, in the sense that the
$J$=2--1 and $J$=3--2 lines are now equally strong. Moreover, the
0.2~M$_\oplus$ model gives CO lines below 0.1~mK, indicating that the
threshold for CO detection occurs between 0.2 and 4~M$_\oplus$ (see
Fig~\ref{co_beta_pic}b, assuming that CO scales linearly in a narrow
range of disk mass) for the distance and inclination of
$\beta$-Pictoris. Roughly a factor of 10 in flux is generally gained
compared to the averaged intensity from the entire disk
(Fig.~\ref{bpic_CO_lines}). A comparison of Fig.~\ref{bpic_CO_lines}d
and Fig.~\ref{co_beta_pic}d reveals that the smaller beams pick up
mostly the warm CO in the inner disk regions.  In addition the
difference in beam dilution is largest for the $J$=2--1 and $J$=3--2
lines compared to the entire disk beam.

On the other hand, there are UV absorption-line studies along the line
of sight towards $\beta$~Pictoris. Roberge et al.~\ (\cite{Roberge})
observed \ion{C}{i} and CO using the HST STIS high-resolution echelle
spectrograph. Contrary to previously detected \ion{C}{i} absorption
(Jolly et al.~\ \cite{Jolly}), an unsaturated spin-forbidden
transition was observed. This allowed for an improved determination of
the ground state C($^3$P) column of (2--4)\,$10^{16}$~cm$^{-2}$.  The
CO absorption results indicate a column of $(6.3 \pm
0.3)$\,$10^{14}$~cm$^{-2}$.

In the 2~M$_\oplus$ disk models, the CO column density is
$8$\,$10^{15}$~cm$^{-2}$, roughly 10 times higher than the observed
value. The \ion{C}{i} column density from the model, $\sim
1$\,$10^{17}$~cm$^{-2}$, is also a factor 2.5--5 higher than observed
in the absorption studies. In case of the 0.2~M$_\oplus$ disk models a
CO and \ion{C}{i} column density of $8$\,$10^{12}$ and
$5$\,$10^{15}$~cm$^{-2}$ is reached respectively. Both are lower than
the observed values, leading to the conclusion that the actual
mass lies in between these models.

The detected [\ion{C}{II}] line provides a reliable estimate
of the amount of ionized gas in the disk around $\beta$-Pictoris.  The
emission line was modeled adopting the ISO beam of 78\,\arcsec. The
integrated emission for the 2 and 0.2~M$_\oplus$ disk models is
$1.9 \times 10^{-21}$ and $2.6 \times 10^{-22}$ W~cm$^{-2}$
respectively. Both are lower than the observed value of $1 \times 10^{-20}$
W~cm$^{-2}$. At first sight, this would indicate that according to
the model more gas is present in the disk around $\beta$-Pictoris.
However, this conclusion depends heavily on the assumed UV radiation
and further modeling is needed.

Freudling et al. \ (\cite{Freudling}) derived from the non-detection
of the \ion{H}{i} 21~cm line an upper limit of
$2-5 \times 10^{19}$~cm$^{-2}$ for the neutral hydrogen column
density. The column calculated for the 2~M$_\oplus$ disk model
including the interstellar field is $8 \times 10^{20}$~cm$^{-2}$ and for
the 0.2~M$_\oplus$ disk model $8 \times 10^{19}$~cm$^{-2}$, a few times
higher than observed.

The \ion{H}{i} column density in the 2M$_{\oplus}$ model is close to
$N({\rm H})=5 \times 10^{20}$~cm$^{-2}$ deduced by Olofsson et al.~\
(\cite{Olofsson}) from their sodium observations using a solar Na/H
abundance ratio.  In the models described here, the available
\ion{H}{i} in the gas is directly related to the amount of H$_{2}$
self-shielding of the stellar radiation.  The interstellar radiation
field penetrating through the disk is not shielded by H$_{2}$, hence
comparable columns of H$_{2}$ and \ion{H}{i} are reached (see paper
I). A disk with a 25 times higher mass would be able to shield the
midplane of the disk and the \ion{H}{i} column will decrease. This
process would hardly affect the carbon species in the outer regions,
because CO will freeze out, incorporating all the available carbon.
In the inner regions, the \ion{C}{i} and CO densities would increase
in a higher mass disk and the column densities would be too high
compared with observations.

\begin{figure}[ht!]
\resizebox{\hsize}{!}{\includegraphics{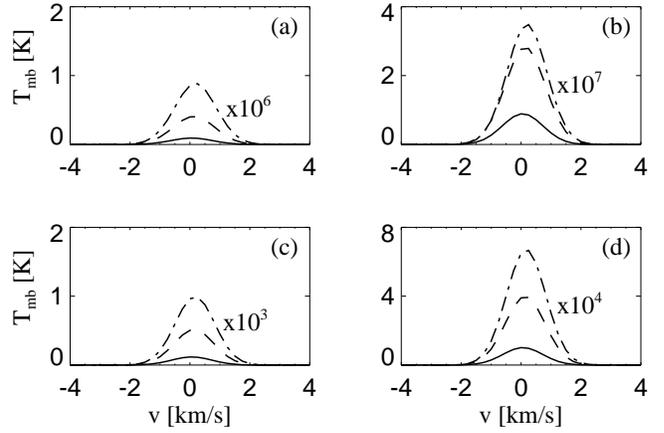}}
\caption{CO rotational line profiles (main beam brightness 
  temperature versus rest velocity) in different Vega models, at
  14.7\,\arcsec\ offset and convolved with the observational beams: (a)
  0.2~M$_\oplus$ model with $v_{\rm drift}=v_{\rm drift}^{\rm max}$,
  (b) 0.2~M$_\oplus$ model with $v_{\rm drift}=0$, (c) 2~M$_\oplus$
  model with $v_{\rm drift}=v_{\rm drift}^{\rm max}$, and (d)
  2~M$_\oplus$ model with $v_{\rm drift}=0$.  The solid line is
  $J$=1--0 (43\,\arcsec ), the dashed line $J$=2--1 (21\,\arcsec ), and
  the dot-dashed line $J$=3--2 (14\,\arcsec ). Note that the
  fluxes are muliplied by the factors indicated.}
\label{co_vega}
\end{figure}

Our models show that a $\beta$-Pictoris disk with 2~M$_\oplus$ of gas
might have escaped detection in CO. The absorption studies towards
$\beta$~Pictoris show that the models give CO and \ion{C}{i} column
densities that are up to a factor of 10 too high. In case of the
0.2~M$_\oplus$ models all of these upper limits are met as C$^{+}$ and
CO$_{\rm ice}$ become the main reservoirs of carbon.  In the case of
the \ion{H}{i} absorption even the low mass 0.2~M$_{\oplus}$ disk
overestimates the observed upper limit. Since the estimate by Olofsson
et al.~(2001) differs by a factor of 10 and the true H$_{2}$ formation
rate used is not known, not too much weight should be given to the
\ion{H}{i} constraint. 

\subsection{Vega}

For Vega, there are in general less observations than for
$\beta$~Pictoris.  Due to the pole-on geometry, absorption line
studies for Vega are impossible. Yamashita et al.~\ (\cite{Yamashita})
searched for the CO $J$=1--0 line with the 45~m Nobayama Radio
Observatory (NRO). They detected no CO emission down to a $T_{\rm
rms}$ of 33~mK in 1~km~s$^{-1}$ bins. Even with the better
sensitivity of $T_{\rm rms}\sim 10$~mK of the JCMT, Dent et al.~\
(\cite{Dent}) did not detect any CO $J$=2--1 around Vega. The above
calculations show that a 2~M$_\oplus$ model for Vega reaches at most
1~mK for the $J$=3--2 transition (Fig.~\ref{co_vega}c). The lowest
$J$=1--0 transition has fluxes below 0.1~mK. This is entirely due to
photodissociation. Simple scaling suggests that the lines will still
not be detectable with current instruments if the disk mass is enlarged 
by a factor of 20.

The difference seen in comparing the models with and without
drift-velocity heating (Fig.~\ref{co_vega}a-d) is entirely due to
differences in gas temperature. Drift-velocity heating is the most
important heating source of the gas and, if omitted, H$_2$
formation/dissociation as well as photoelectric heating of the
$\mu$m-size dust particles remain, leading to lower gas
temperatures. The CO fluxes are a factor of 10 higher than with the
beam of 128.9\,\arcsec\ corresponding to the entire disk (compare
Fig.~\ref{vega_CO_overview} and \ref{co_vega}). This is due to the low
CO densities in the outer regions picked up in the large beam. The
sometimes different line ratios can be explained by the restricted
area picked up by the smaller telescope beams.

The models for the Vega disk predict [\ion{C}{ii}] integrated
emission lines of 1.9$10^{-20}$ and 7.9$10^{-22}$~W~cm$^{-2}$ for
the 2 and 0.2~M$_\oplus$ disks. The former integrated intensity corresponds
to a peak intensity of $3.4 \times 10^{-20}$~W~cm$^{-2}~\mu$m$^{-1}$ 
and the latter to $1.4 \times 10^{-21}$~W~cm$^{-2}~\mu$m$^{-1}$.  
The ISO satellite observed Vega in the wavelength range of the 
[\ion{C}{ii}] 157.74~$\mu$m line (Sect.~\ref{cii_obs}) and obtained 
an upper limit of 
$1.6 \times 10^{-21}$~W~cm$^{-2}~\mu$m$^{-1}$. This would suggest a
maximum amount of gas of $\sim 0.2~M_\oplus$ in the disk around Vega.

\section{Future observations}

We present here [\ion{C}{i}] predictions from our models appropriate
for the new MPIfR/SRON 800 GHz and 460 GHz single channel receivers to 
be installed at the
Atacama Pathfinder Experiment (APEX) in 2003. APEX is located at 5000
m at Chajnantor in Chili and is perfectly placed to observe
$\beta$-Pictoris. For Vega, the [\ion{C}{i}] emission line is
calculated using the JCMT telescope instead of APEX. Additionally
[\ion{C}{i}] and [\ion{C}{ii}] lines for the future instrumentation on
board of SOFIA and Herschel are calculated.  The velocity resolution
of all the telescopes is set at 1 km s$^{-1}$ to detect the line
profiles except for the [\ion{C}{i}] 809.3~GHz (370~$\mu$m) line, where 
5~km~s$^{-1}$ is used.

\begin{figure}[ht!]
\resizebox{\hsize}{!}{\includegraphics{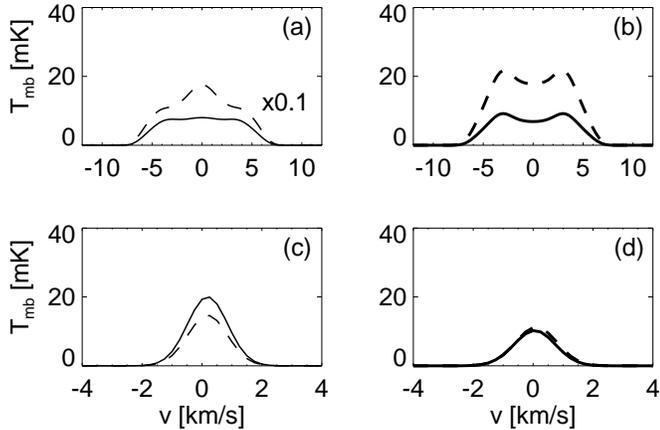}}
\caption{[\ion{C}{i}] fine structure line profiles (main beam brightness 
  temperature versus rest velocity): (a) 2~M$_\oplus$ model for
  $\beta$~Pictoris with $v_{\rm drift}=0$ and interstellar radiation
  field appropriate for the APEX (note that the flux is muliplied by
  0.1), (b) same as (a) but appopriate for Herschel, (c) 2~M$_\oplus$
  model for Vega with $v_{\rm drift}=0$ appropriate for JCMT, (d) same
  as (c) but appropriate for Herschel. The solid line denotes the
  $^{3}$P$_{1}$--$^{3}$P$_{0}$ transition at 609.13~$\mu$m (10, 13,
  and 44\,\arcsec\ beam for JCMT, APEX, and Herschel respectively) and
  the dashed line the $^{3}$P$_{2}$--$^{3}$P$_{1}$ transition at
  370.42~$\mu$m (6, 8, and 26\,\arcsec\ beam for JCMT, APEX and
  Herschel, respectively)}
\label{c_bpicandvega}
\end{figure}

Since detailed sensitivity estimates are not yet available for APEX,
the limits in the appropriate bands are calculated for the JCMT. 
In general, APEX, due to its excellent site and stable
atmosphere has better sensitivities.

 The [\ion{C}{i}] 370.42~$\mu$m intensity of the 2~M$_\oplus$ disk
model with interstellar UV field for $\beta$~Pictoris yields a flux of
300~mK, whereas the 2~M$_\oplus$ Vega model yields only 15~mK. Since
the source is at a low zenith angle, the current 370~$\mu$m receiver at
the JCMT would require 200 minutes of observing time for a 5
km~s$^{-1}$ velocity resolution to reach a 3\,$\sigma$ detection for a
300~mK peak intensity of the line. This is clearly not feasible. The
expected upgraded receiver at APEX would reach this limit in about 20
minutes for $\beta$-Pictoris. A higher spectral resolution of 
1~km~s$^{-1}$ would increase the time to about 8 hours.

The lowest [\ion{C}{i}] line at 492.2~GHz (609.13~$\mu$m) is calculated using the
JCMT specifications.  A typical noise level of $T_{A}^{\ast}$=40~mK in
1~km~s$^{-1}$ bin could be reached in 1.1 hours of integration time at
an elevation of 60\degr, appropriate for the $\beta$-Pictoris disk
from the APEX site. The predicted $T_{mb}=100$~mK line should be
detectable with this configuration. For a star like Vega with a strong
UV radiation field, this line is not observable in low mass
($\le 2$~M$_\oplus$) disks.

For the HIFI instrument on board of the Herschel satellite, the
[\ion{C}{i}] line at 609.13~$\mu$m is at the lower limit of Band 1,
while the 370.42~$\mu$m line falls in Band 3.  The sensitivity
estimates for a long integration (5 hours) are 2 and 3~mK respectively
(de Graauw \& Helmich \cite{Graauw}). Since the Herschel beam is much
larger than the JCMT beam, nearby edge-on disks will suffer from large
beam dilution. Still, both the edge-on $\beta$-Pictoris disk and the
pole-on disk around Vega should be just observable in both lines.

\begin{figure}[ht!]
\resizebox{\hsize}{!}{\includegraphics{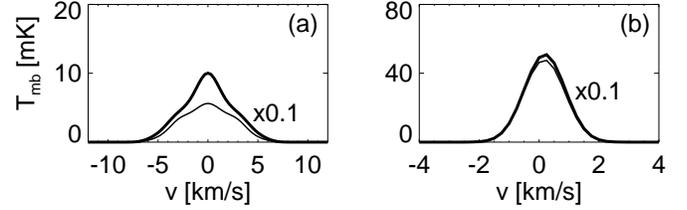}}
\caption{[\ion{C}{ii}]\,157.74~$\mu$m fine structure line profiles 
  (main beam brightness temperature versus rest velocity): (a)
  2~M$_\oplus$ model for $\beta$~Pictoris with $v_{\rm drift}=0$ and 
  interstellar radiation field, (b) 2~M$_\oplus$ model for Vega with 
  $v_{\rm drift}=0$. The thick line denotes the Herschel beam of
  11\,\arcsec, the thin line the SOFIA beam of 16\,\arcsec. Note that the
  fluxes are muliplied by the factors indicated.}
\label{c+_bpicandvega}
\end{figure}

The [\ion{C}{ii}] fine structure line at 157.74~$\mu$m can be observed
in Band 6 of the HIFI instrument on board of Herschel.  In a long
exposure (5 hours), a sensitivity of $\sim$6~mK is reached (de Graauw
\& Helmich \cite{Graauw}). A 2~M$_\oplus$ model will be observable in
both Vega and $\beta$-Pictoris.

If the slightly larger beam of SOFIA is used, 16\,\arcsec\ com\-pared
to the 11\,\arcsec\ of Herschel, the model fluxes for the
[\ion{C}{ii}] line decrease by about a factor 2 for an edge-on disk
model due to additional beam dilution. Since we applied an offset of
15\,\arcsec\ to the pole-on disk model, there is hardly any difference
between the two beam sizes for this model.  In both cases, the beam is
entirely filled by the disk model. Apart from the difference in beam size,
HIFI and the GREAT instrument on SOFIA will have comparable sensitivity
limits and integration times.

The Atacama Large Millimeter Array (ALMA) will reach a
sensitivity and angular resolution which will be orders of magnitude
larger than the current status. In the case of Vega and
$\beta$~Pictoris, angular resolution is of minor importance due to the
relatively small stellar distances. Adopting the expected
sensitivities, an hour of integration on the CO 3--2 line with a
spectral resolution of 0.5~km~s$^{-1}$ would give a 5\,$\sigma$
detection of 0.05~K. This result is obtained for the compact configuration
with baselines up to 150~m, which gives a spatial
resolution of 2\,\arcsec.  Figure \ref{bm2.0_hclp_IS_integrated} shows
that the integrated line intensity of the CO 3--2 line peaks at a
2\,\arcsec\ offset for $\beta$-Pictoris. Using a spectral resolution of
0.5~km~s$^{-1}$ the local emission at 2\,\arcsec\ offset reaches
2.1~K. At the wavelength of the [\ion{C}{i}] $^{3}P_{1}-^{3}P_{0}$
transition the 5\,$\sigma$ level reached with ALMA is expected to be
twice as high (0.1~K). The modeled intensity for this line peaks at
1.2~K. In both cases the CO and [\ion{C}{i}] line should be detectable
in a few seconds and half a minute respectively. For Vega the
detection limits are more stringent and peak intensities (with a
2\,\arcsec\ beam) will be 0.05 for the [\ion{C}{i}] line and
$6 \times 10^{-4}$~K for the CO line. [\ion{C}{i}] will thus be
detectable in 4 hours integration time and CO is undetectable (7000
hrs) for this binsize at a 5\,$\sigma$ level.

\section{Discussion and conclusion}

The optically thin models described in this paper provide
a tool to constrain the gas mass in circumstellar disks on
the basis of observed emission lines and derived column
densities. Due to the complex interplay between chemistry,
temperature, shielding, and radiative transfer in the emitting
lines, it cannot be exluded that more
massive optically thick disk models yield similar column
densities and line emission, but this is left for future
modeling.

Assuming that CO emission scales linearly over a small range of disk
mass, the comparison with observations of CO and [\ion{C}{I}] 
made for $\beta$~Pictoris constrains the gas mass between 0.2 and
4~M$_\oplus$. The question remains: do the results indeed lead to the
conclusion that the $\beta$~Pictoris disk is depleted in gas, or is
there still some physical input missing in the models?

Evidence that our models are still not complete comes from the recent
observations of Lecavelier des Etangs et al.~\ (\cite{Lecavelier}) and Deleuil et
al.\ (\cite{Deleuil}), who have shown that $\beta$~Pictoris might be
an active star. The detection of the \ion{O}{vi} emission doublet at
1035~\AA\, and of \ion{C}{ii} and \ion{C}{iii} lines with the FUSE
satellite, lead to the conclusion that $\beta$~Pictoris may have a
chromosphere. Even though the observed \ion{O}{vi} lines do not
overlap with any of the CO photodissociation lines, the continuum of
the chromosphere increases the stellar UV flux and hence the amount of
CO photodissociation. H$_2$ will be less affected due to its more
efficient self-shielding. The existence of a chromosphere in
$\beta$~Pictoris would increase the photodissociation rate for CO as
well as the ionization rate for carbon, hence leading to lower CO and
\ion{C}{i} column densities.

FUSE is not sensitive enough to detect the continuum flux shortward of
1100~\AA\ in $\beta$~Pictoris. This makes it difficult to estimate the
strength of the additional chromospheric UV radiation field.
Nevertheless, the upper limit for a continuum given by the FUSE
spectra (at 1000~\AA\ $\sim 3.2\times
10^{-11}$~erg~cm$^{-2}$~s$^{-1}$~Hz$^{-1}$), could still hide a
chromosphere of about 10 times higher than the photospheric flux
included so far in the models. Section~\ref{results} shows that the
inclusion of an interstellar UV radiation field decreases the CO
emission already by a factor of 20 (compare Fig.~\ref{co_beta_pic}a
and c). Hence the presence of a chromosphere in the star
$\beta$~Pictoris, which can be according to the FUSE data up to a
factor 10 in units of the interstellar radiation field, may allow
larger disk masses to be consistent with the current CO $J$=2--1 and 3--2 
observational limits. Moreover it can enlarge the [\ion{C}{ii}] 
157.74~$\mu$m line significantly and thus bring it closer to the 
observed integrated intensity.

Heap et al. (\cite{Heap:00}) report the detection of an inner warp
in the disk around $\beta$~Pictoris. From STIS coronographic observations,
they deduce a two component disk model: a main outer disk and a fainter
inner disk, which is inclined by 4-5\degr\ with respect to the main disk
and extends to about 80~AU. Hence, the inclined inner disk will receive
more stellar UV radiation, leading to less CO, \ion{C}{i} and more
\ion{C}{ii}. In the case of an inner warp, the UV absorption-line studies 
would miss some of the inner disk material and underestimate the total 
disk mass. Both effects allow larger disk masses to be consistent with the
present observations.

So we are left with the final question: do the dusty disks seen around
A~stars (``Vega-type'' stars) still contain gas? The previously drawn
conclusion that these disks are debris disks which contain less gas
than dust, is mainly based on CO conversion factors appropriate for
molecular clouds. But the low mass disks around A-type stars resemble
more a photodissociation region than a molecular cloud. From the
models presented here and the discussion of the various observations
for $\beta$~Pictoris, it is difficult to keep the gas-to-dust ratio as
high as 100. On the basis of the present models, we argue for a 
gas-to-dust ratio between 0.5 and 9 for $\beta$~Pictoris, assuming 
a dust mass of 0.44~M$_\oplus$ (Chini et al. \cite{Chini:91}). This ratio depends
of course on the dust mass and on the basis of different modeling approaches 
Dent et al. (\cite{Dent:00}) found a dust mass of 0.04~M$_\oplus$,
while Li \& Greenberg (\cite{Li:98}) obtained 0.33~M$_\oplus$. Any chromosphere 
included in the models will --- depending on its strength --- raise 
the gas-to-dust value. Moreover, the 2~M$_\oplus$ model with the 
interstellar radiation field suggests that the inclusion of a 
chromospheric UV radiation field can solve the remaining problem 
with CO, \ion{C}{i}, and \ion{C}{ii} simultaneously. This shows 
the power of using observations 
of several species to constrain the gas mass instead of using only one, 
namely CO.

For Vega, the non-detection of CO gave an upper limit for the
gas mass of $7\times 10^{-3}$~M$_\oplus$ using a CO abundance
of $\sim 10^{-4}$ (Yamashita et al. \cite{Yamashita}). Our models show, 
that CO is entirely photodissociated in Vega and hence this upper 
limit has to be regarded with caution. Following the models presented
in this paper, most of the carbon is in form of C$^+$. Hence,
we use [\ion{C}{ii}] observations to constrain the gas mass in the
disk around Vega and obtain an upper limit of 0.2~M$_\oplus$. This
gives and upper limit of 33 for the gas-to-dust ratio, assuming a 
dust mass of $6\times 10^{-3}$~M$_\oplus$ (Chini et al. \cite{Chini:90})
and 80 assuming a dust mass of $2.5\times 10^{-3}$~M$_\oplus$  
(Dent et al. \cite{Dent:00}).

In order to improve our understanding of these disks and to constrain
the gas mass, more suitable gas tracers than CO are needed. While
absorption studies only probe the disk material in the line of sight,
emission line studies contain information on the disk structure as a
whole and are therefore better suited to constrain the disk
models. The calculations presented in this paper for a 2~M$_\oplus$
disk with a gas-to-dust mass ratio of 100 show that the best tracers
among those considered here (CO, C, and C$^+$) are C$^+$ and C. This
result does not depend on the geometry of a nearby disk, because the
beams of APEX, SOFIA and Herschel are small.  Moreover the conclusion
holds for disks around A stars with a moderate radiation field like
$\beta$~Pictoris as well as for disks that are exposed to a larger UV
flux like around Vega.

\begin{acknowledgements}
The authors are grateful to M. Hogerheijde and F. van der Tak for 
use of their 2D Monte Carlo code. The JCMT data have been obtained 
from the Canadian Astronomy Data Center, which is operated by the 
Dominion Astrophysical
Observatory for the National Research Council of Canada's Herzberg
Institute of Astrophysics. I.~Kamp acknowledges support by a Marie
Curie Fellowship of the European Community programme ``Improving Human
Potential'' under contract number MCFI-1999-00734.  Astrochemistry in
Leiden is supported by a SPINOZA grant from the Netherlands
Organization for Scientific Research (NWO).
\end{acknowledgements}


\begin{thebibliography}{99}
 \bibitem[2001]{Augereau}
  Augereau, J. C., Lagrange, A. M., Mouillet, D., \& M\'{e}nard, F. 2001, A\&A, 365, 78
 \bibitem[1999]{Barrado}
  Barrado y Navascu\'{e}s, D., Stauffer, J. R., Song, I., \& Caillault, J.-P. 1999, ApJ, 520, L123
 \bibitem[1992]{cheng} 
  Cheng, K.-P., Bruhweiler, F. C., Kondo, Y., \& Grady, C. A. 1992, \apjl, 396, L83 
 \bibitem[1990]{Chini:90} 
  Chini, R., Kruegel, E. \& Kreysa, E. 1990, A\&A, 227, L5
 \bibitem[1991]{Chini:91} 
  Chini, R., Kruegel, E., Kreysa, E., Shustov, B., \& Tutukov, A. 1991, \aap, 252, 220
 \bibitem[1975]{Chu}
  Chu, S.-I., \& Dalgarno, A. 1975, Royal Society (London), Proceedings, Series A, vol.342, no.1629, 191
 \bibitem[2001]{Graauw}
  de Graauw, Th., \& Helmich, F. P. 2001, in ESA SP-460, eds. G. L. Pilbratt, J. Cernicharo, 
     A. M. Heras, T. Prusti, R. Harris, 45
 \bibitem[2001]{Deleuil} 
  Deleuil, M., Bouret, J.-C., Lecavelier des Etangs, A., et al.\ 2001, \apjl, 557, L67
 \bibitem[1995]{Dent}
  Dent, W. R. F., Greaves, J. S., Mannings, V., Coulson, I. M., \& Walther, D. M.,
  1995, MNRAS, 277, L25
 \bibitem[2000]{Dent:00}
  Dent, W. R. F., Walker, H. J., Holland, W. S., Greaves, J. S. 2000, MNRAS, 314, 702
 \bibitem[1996]{Dutrey}
  Dutrey, A., Guilloteau, S., Duvert, G., et al.\ 1996, A\&A, 309, 493
 \bibitem[1977]{Flower}
  Flower, D. R., \& Launay, J. M. 1977, J.\ Phys.\ B, 10, 3673
 \bibitem[1995]{Freudling} 
  Freudling, W., Lagrange, A.-M., Vidal-Madjar, A., Ferlet, R., \& Forveille, T. 
  1995, \aap, 301, 231
 \bibitem[1968]{Habing} 
  Habing, H. 1968, Bull.\ Astr.\ Inst.\ Netherlands, 19, 421
 \bibitem[2001]{habingb}
  Habing, H. J., Dominik, C., Jourdain de Muizon, M., et al. 2001, A\&A, 365, 545
 \bibitem[1985]{Hayashi}
  Hayashi, C., Nakazawa, K., \& Nakagawa, Y. 1985, in  Protostars \& Planets II,
      eds. D. C. Black, M. S. Mathews, 1100
 \bibitem[2000]{Heap:00}
  Heap, S. R., Lindler, D. J., Lanz, T. M., et al.\ 2000, ApJ, 539, 435
 \bibitem[2000]{HvdT00}
  Hogerheijde, M. R., \& van der Tak, F. 2000, A\&A, 362, 697
 \bibitem[1998]{Holland:1998}
  Holland, W. S., Greaves, J. S., Zuckerman, B., et al. 1998, Nature, 392, 788
 \bibitem[1998]{Jolly} 
  Jolly, A., Mc Phate, J. B., Lecavelier, A., et al.\ 1998, \aap, 329, 1028
 \bibitem[2000]{Kamp1}
  Kamp, I., \& Bertoldi, F. 2000, A\&A, 353, 276 (Paper\,{\sc i})
 \bibitem[2001]{Kamp2}
  Kamp, I., \& van Zadelhoff, G.-J. 2001, A\&A, 373, 641 (Paper\,{\sc ii})
 \bibitem[1986]{Keenan}
  Keenan, F. P., Lennon, D. J., Johnson, C. T., \& Kingston, A. E. 1986, MNRAS, 220, 571
 \bibitem[1992]{Kurucz}
  Kurucz, R. L. 1992, Rev.\ Mex.\ Astron.\ Astrofis., 23, 181
 \bibitem[1977]{Launay}
  Launay, J. M., \& Roueff, E. 1977, A\&A, 56, 289
 \bibitem[2001]{Lecavelier}
  Lecavelier des Etangs, A., Vidal-Madjar, A., Roberge, A. et al.\ 2001, Nature, 412, 706
 \bibitem[1998]{Li:98}
  Li, A., Greenberg, J. M. 1998, A\&A, 331, 291
 \bibitem[1998]{Liseau1} 
  Liseau, R., \& Artymowicz, P. 1998, \aap, 334, 935
 \bibitem[1999]{Liseau2} 
  Liseau, R. 1999, \aap, 348, 133
 \bibitem[1983]{Mendoza}
  Mendoza, C. 1983, in IAU Symposium 103, Planetary nebulae, ed. D. R. Flower, 143
 \bibitem[2001]{Olofsson}
  Olofsson, G., Liseau, R., \& Brandeker, A. 2001, ApJL, 563, L77 
 \bibitem[2000]{Roberge}
  Roberge, A., Feldman, P. D., Lagrange, A. M., et al.\ 2000, \apj, 538, 904 
 \bibitem[1994]{Savoldini} 
  Savoldini, M., \& Galletta, G. 1994, A\&A, 285, 467
 \bibitem[1985]{Schinke}
  Schinke, R., Engel, V., Buck, U., Meyer, H., \& Diercksen, G. H. F. 1985, ApJ, 299, 939
 \bibitem[1991]{Schroeder}
  Schr\"{o}der, K., Staemmler, V., Smith, M. D., Flower, D. R., \& Jacquet, R. 1991, J.\ Phys.\ B, 
    24, 2487
 \bibitem[1995]{Stauffer}
  Stauffer, J. R., Hartmann, L. W., \& Barrado y Navascu\'{e}s, D. 1995, ApJ, 454, 910
 \bibitem[1996]{Swinyard}
  Swinyard, B. M., Clegg, P. E., Ade, P. A. R., et al. 1996, A\&A, 315, L43
 \bibitem[2001]{Thi}
  Thi, W. F., Blake, G. A., van Dishoeck, E. F., et al.\ 2001, Nature, 409, 60 
 \bibitem[2001b]{Thib} 
  Thi, W. F., van Dishoeck, E. F., Blake, G. A., et al.\ 2001b, \apj, 561, 1074 
 \bibitem[1993]{Yamashita}
  Yamashita, T., Handa, T., Omodaka, T., et al.\ 1993, ApJ, 402, L65
 \bibitem[1995]{Zuckerman}
  Zuckerman, B., Forveille, T., \& Kastner, J. H. 1995, Nature, 373, 494
 \end{thebibliography}
\end{document}